\documentclass{iucrjournals}

\usepackage{wrapfig}
\usepackage[nolist]{acronym}

\begin{acronym}
\acro{DLSR}{Diffraction-Limited Storage Ring}
\acro{XMPI}{X-ray Multi-Projection Imaging}
\acro{API}{Application Programming Interface}
\acro{FOV}{field-of-view}
\end{acronym}

\title{Time-resolved 3D imaging opportunities with XMPI at ForMAX}

\author[a]{Julia Katharina Rogalinski\IUCrCemaillink{julia_katharina.rogalinski@fysik.lu.se}\IUCrOrcidlink{0009-0008-7133-619X}}
\author[a]{Zisheng Yao\IUCrOrcidlink{0000-0001-7970-2133}}
\author[a]{Yuhe Zhang\IUCrOrcidlink{0000-0003-2718-6434}}
\author[a]{Zhe Hu\IUCrOrcidlink{0009-0000-1329-1634}}
\author[b,c]{Korneliya Gordeyeva\IUCrOrcidlink{0000-0002-1195-1405}}
\author[b,c]{Tomas Rosén}
\author[b,c]{Daniel Söderberg\IUCrOrcidlink{0000-0003-3737-0091}}
\author[d]{Andrea Mazzolari}
\author[e]{Jackson da Silva} 
\author[e]{Vahid Haghighat} 
\author[e]{Samuel A. McDonald}
\author[e]{Kim Nyg{\aa}rd\IUCrOrcidlink{0000-0002-4906-0093}}
\author[a,e]{Eleni Myrto Asimakopoulou\IUCrOrcidlink{0000-0003-2127-373X}}
\author[a]{Pablo Villanueva-Perez\IUCrOrcidlink{0000-0002-4671-790X}}

\affil[a]{Synchrotron Radiation Research and NanoLund, Lund University, Lund, Sweden}
\affil[b]{Department of Fibre and Polymer Technology, Royal Institute of Technology, Stockholm, Sweden}
\affil[c]{Wallenberg Wood Science Center, Royal Institute of Technology, Stockholm, Sweden}
\affil[d]{INFN Section of Ferrara, Ferrara, Italy}
\affil[e]{MAX IV Laboratory, Lund University, Lund, Sweden}

\begin{document} 
\maketitle 

\begin{abstract}
X-rays are commonly used in imaging experiments due to their penetration power, which enables non-destructive resolution of internal structures in samples that are opaque to visible light. Time-resolved X-ray tomography is the state-of-the-art method for obtaining volumetric 4D (3D + time) information by rotating the sample and acquiring projections from different angular viewpoints over time. This method enables studies to address a plethora of research questions across various scientific disciplines. 
However, it faces several limitations, such as incompatibility with single-shot experiments, challenges in rotating complex sample environments that restrict the achievable rotation speed or range, and the introduction of centrifugal forces that can affect the sample's dynamics. 
These limitations can hinder and even preclude the study of certain dynamics.
Here, we present an implementation of an alternative approach, X-ray Multi-Projection Imaging (XMPI), which eliminates the need for sample rotation. Instead, the direct incident X-ray beam is split into beamlets using beam splitting X-ray optics. 
These beamlets intersect at the sample position from different angular viewpoints, allowing multiple projections to be acquired simultaneously. 
We commissioned this setup at the ForMAX beamline at MAX~IV, the first operational Diffraction-Limited Storage Ring. 
We present projections acquired from two different sample systems - fibers under mechanical load and particle suspension in multi-phase flow - with distinct spatial and temporal resolution requirements.
We demonstrate the capabilities of the ForMAX XMPI setup using the detector's full ADC range for the relevant sample-driven spatiotemporal resolutions: i) at least 12.5~kHz framerates with 4~\textmu m pixel sizes (fibers) and ii) 40 Hz acquisitions with 1.3~\textmu m pixel sizes (multi-phase flows).
The presented setup and results set the basis for a permanent XMPI endstation at ForMAX, enabling studies of samples that cannot be studied with state-of-the-art time-resolved tomography and offering flexibility to adapt to the spatiotemporal requirements of the studied dynamics.
\end{abstract}

\keywords{time-resolved 3D imaging; X-ray imaging, X-ray multi-projection imaging; MAX~IV; ForMAX beamline.}

\section{Introduction}
Advancements in accelerator technologies have prompted the development of \acp{DLSR}, also known as 4th generation synchrotron light sources. Using multibend achromats, the emittance of the electron beam can be significantly decreased, leading to an increase in brilliance by up to several orders of magnitude compared to their predecessors \cite{eriksson2014DLSR, raimondi2023DLSR}. As a consequence, the flux density, i.e., the number of photons per unit of area and time, is increased, thereby creating new opportunities to achieve higher spatiotemporal resolution in X-ray imaging experiments \cite{villanova2017fast,XrayTomographyTomoscopyReview, yao2024DLSR}.

The MAX~IV Laboratory in Lund, Sweden, is the first operational \ac{DLSR} worldwide, providing beamlines with unprecedented flux capabilities. The ForMAX beamline, designed for scattering and imaging experiments, delivers synchrotron radiation through an undulator. A full undulator harmonic yields high spectral flux density, i.e., number of photons per unit of time per unit of area per bandwidth, within a well-defined energy bandwidth, in contrast to broader spectra generated, e.g., by wigglers. The narrow energy bandwidth enables quantitative analysis of material density and composition, avoiding complications introduced by polychromatic beams \cite{Maire2014}. ForMAX delivers an exceptionally high photon flux of approximately $5\cdot10^{14} \frac{\textrm{ph}}{\textrm{s}}$ \cite{nygaard2024formax}, and consequently high photon flux density. This capability supports highly time-resolved imaging experiments, including time-resolved tomography, the state-of-the-art method for acquiring time-resolved volumetric information in a non-destructive manner \cite{yao2024DLSR}. 

Synchrotron X-ray tomography is performed by rotating a sample relative to an incident beam, thereby acquiring projections (tomograms) of the sample.
A detector positioned downstream of the sample collects a sequence of tomograms, typically over a $180^\circ$ rotation, which is used to reconstruct the full sample volume.
Time-resolved tomography requires the acquisition of a complete tomographic dataset within the temporal resolution of interest. 
For example, to resolve $10~\textrm{Hz}$ dynamics, a sample rotation with a speed of at least $10~\textrm{revolutions-per-second}$ ($10~\textrm{Hz}$) is required, assuming one complete 3D dataset every $180^\circ$ and Nyquist sampling.
The higher the temporal resolution, the more stringent the requirements become for rotation stages, detector efficiency, photon flux, and the design for in situ and operando sample environments.
While technological advancements have improved detectors and X-ray sources, achieving stable high-speed sample rotation remains a significant challenge. 
Time-resolved tomography has been demonstrated at temporal resolutions up to $1~\textrm{kHz}$ with a spatial resolution of 8.2~\textmu m, using a highly customized experimental setup to study the evolution of metallic foams \cite{Moreno:2021}. 
Such advancements, while impactful, underscore the inherent limitations of this technique for time-resolved volumetric imaging of multiple sample categories, due to the fundamental requirement for sample rotation. 
Rotation-sensitive samples, such as dense liquids and soft materials, have intrinsic dynamics that can be affected by the centrifugal forces generated during rotation, thereby compromising the observations. Moreover, samples requiring complex experimental environments face significant challenges with mechanical stability, cable management, and uniform sample visibility, which often affects the quality of the collected data. Samples with above $20~\textrm{Hz}$ dynamics often require custom rotation stages, which are not widely accessible, cannot serve as a universal solution for all samples, and may struggle to exceed $1000~\textrm{revolutions-per-second}$~\cite{XrayTomographyTomoscopyReview}. 
Additionally, single-shot phenomena, such as shock wave propagation \cite{schropp2015imaging} or investigations of defect effects \cite{kumar2016strength}, cannot be captured with conventional tomography on a single sample. These studies rely instead on the possibility of preparing identical sample copies and reproducible processes, which is not always possible.
As a result, time-resolved tomography leaves a vast spatiotemporal domain unexplored.


These limitations hinder the progress of numerous scientific and industrial studies.
Some indicative examples are studies of (i) multi-phase opaque flows through complex geometries (e.g., blood flow in veins), where radial forces interfere with natural flow behavior, (ii) mechanical properties of novel, sustainable materials (e.g., cellulose foams and wood fibers), which are too delicate and prone to damage under rotation, and (iii) damage mechanisms on materials such as fiber-reinforced polymer composites, which are widely used in engineering applications, where failure modes occur within milliseconds, or less, and require temporal resolutions exceeding $1~\textrm{kHz}$.

Alternative methods for acquiring time-resolved volumetric information are required to meet the growing needs of the scientific community.
The approach that is followed here is \ac{XMPI} \cite{Hoshino2013, Villanueva-Perez:18, duarte2019computed, Bellucci:23, voegeli2023, asimakopoulou2024xmpi, voegeli2024multibeam, Yahsiro10676953, Sumiishi10676491}, a technique that provides multiple, angularly spaced illumination points on the sample position by employing beam splitting optics, enabling volumetric data acquisition without the need for sample rotation.
By eliminating the requirement for rotation, \ac{XMPI} opens new possibilities for studying complex sample environments and sensitive dynamic processes.
The employment of \ac{XMPI} at high-spectral flux density facilities such as ForMAX (MAX~IV) enables access to previously inaccessible temporal resolutions in systems that are incompatible with tomography, thus allowing the exploration of new physical phenomena.

\ac{XMPI} has become feasible through advances in fast detector readout schemes, improvements in crystal manufacturing for beam splitting optics, and the development of novel reconstruction algorithms. There remains significant potential to further optimize XMPI by exploring alternative crystal materials with enhanced diffraction efficiency, testing different beam splitting schemes to explore various geometries, and implementing improved mounting and cooling strategies to reduce artifacts caused by heat load~\cite{bellucci2024crystals}. In parallel, ongoing developments on reconstruction algorithms aim to improve reconstruction quality when only a sparse number of projections is available~\cite{zhang20244donix, hu2025super}.

The paper introduces an \ac{XMPI} setup tailored to the ForMAX beamline at MAX~IV and demonstrates its application in two scientific applications enabled by \ac{XMPI}: (i) the loading behavior of wood fibers at different stages of Kraft cooking and (ii) particle suspension in multi-phase flows.
The structure of the paper is as follows.
First, we introduce in Section~\ref{sec:Instrumentation} the ForMAX beamline, the \ac{XMPI} instrumentation, and the data processing.
In Section~\ref{sec:ScientificApplications}, we discuss the studied sample processes, motivate the need to address them with \ac{XMPI}, and discuss details of the experimental apparatus. 
In Section~\ref{sec:Discussion}, we discuss our results.
Finally, in Section~\ref{sec:ConclusionAndOutlook}, we conclude our results, provide an outlook for future improvements, and highlight additional scientific applications that could benefit from \ac{XMPI} at ForMAX.

\section{Instrumentation}
\label{sec:Instrumentation}
\subsection{The ForMAX Beamline}
The ForMAX beamline of MAX~IV \cite{nygaard2024formax} was developed to meet the need for structural characterization of hierarchical materials. 
Its technical design has been tailored to offer complementary experimental modalities, enabling the investigation of materials across multiple length scales - from the nanometer scale using small- and wide-angle X-ray scattering (SWAXS) to the micro- and millimeter scale using full-field synchrotron X-ray microtomography (SRµCT). 
The beamline is equipped with a $3~\textrm{m}$ long room-temperature in-vacuum undulator, with a period length of $17~\textrm{mm}$ and maximum effective deflection parameter of $\textrm{K}=1.89$ at the minimum magnetic gap of $4.5~\textrm{mm}$. 
The fifth to thirteenth harmonics of the undulator are used to operate within the energy range of $8\textrm{-}25~\textrm{keV}$. 
The harmonic peaks exhibit a narrow energy profile ($\Delta\textrm{E} \approx 100~\textrm{eV}$), which facilitates quantitative analysis of the collected data.

ForMAX can operate with either a double-crystal monochromator (DCM) or a double-multilayer monochromator (MLM), depending on the needs of the experiment.
Photon-demanding experiments, such as time-resolved experiments, benefit from the higher photon flux provided by the MLM, due to its larger bandpass ($\Delta\textrm{E}/\textrm{E}\simeq 1\%$). This configuration yields a beam size of $1.3~\textrm{mm} \times 1.5~\textrm{mm}$ in the sample position, with a measured flux of $10^{14}\textrm{ - }5\cdot10^{14} \frac{\textrm{ph}}{\textrm{s}}$ in the energy range of $9-20~\textrm{keV}$ at the minimum undulator gap. 
The photon flux is distributed over a narrow energy range, resulting in a high spectral flux density.
As an example, at $16.5~\textrm{keV}$ with the smallest insertion-device gap, the photon flux distribution has a full-width-half-maximum of $100~\textrm{eV}$. 
This high spectral density makes ForMAX an ideal environment for deploying \ac{XMPI}, as discussed in the following section, and enables the study of phenomena that remain inaccessible with conventional time-resolved tomography. 

\subsection{The XMPI Setup}
\begin{figure}
    \centering
    \includegraphics[width=\textwidth]{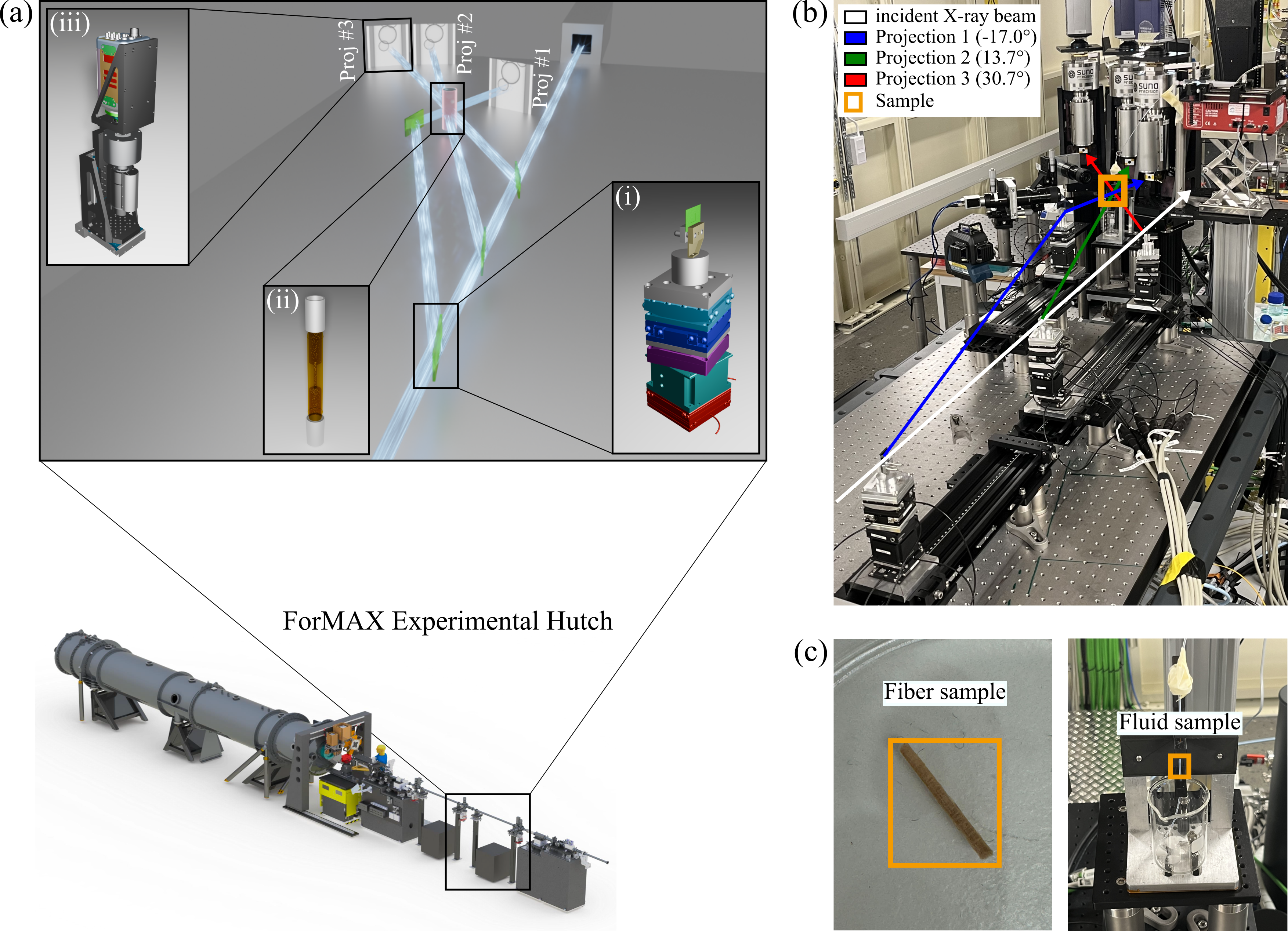}
    \caption{The \ac{XMPI} setup at ForMAX. (a) The \ac{XMPI} setup consists of beam splitters mounted on nanopositioners (i) for the generation of beamlets through spectral and amplitude splitting, that intersect at the sample environment position (ii), and an indirect detector system for each projection (iii). This illustration showcases an example study where the dynamics of interest concern the flow of particles through a capillary. The setup was commissioned upstream of the main sample station of ForMAX by modifying the box-marked segment in the right-hand side beamline sketch. (b) A photograph of the setup in the ForMAX beamline with annotated projections. (c) Examples of samples that can be examined using \ac{XMPI}: Fiber (left) and fluid samples (right).}
    \label{fig:XMPI_setup}
\end{figure}

The \ac{XMPI} setup consists of three major components, depicted in Figure~\ref{fig:XMPI_setup}(a): (i) beam splitters for the generation of beamlets, i.e., sub-fractions of the direct incident beam that travel in air and intersect at the (ii) sample environment position, providing simultaneous, angularly-resolved illumination, and (iii) an indirect detector system for each projection \cite{asimakopoulou2024xmpi}. 
The splitting setup is inspired by interferometers, split-and-delay lines, and specific setups for single-shot imaging as shown in References~\cite{Oberta2013_Splitters,Mokso2015_StereoDualEnergy}. The setup was commissioned at a dedicated section of the beamline, located upstream of the main endstation. The setup was installed by removing the vacuum pipes and installing two optical tables; one hosting the beam splitters and one for the sample environment and detectors, as illustrated in Figure~\ref{fig:XMPI_setup}(b). Two examples of samples that can be examined using \ac{XMPI} are shown in Figure~\ref{fig:XMPI_setup}(c).

\subsubsection{Crystals}
The XMPI setup aims to generate multiple illumination viewpoints on the sample position. This is achieved by employing perfect crystals for the generation of beamlets from an incident X-ray beam via two different splitting schemes: spectral and amplitude splitting. The former refers to the redirection of a portion of the incident beam's spectrum, which has a bandwidth $\Delta\textrm{E}$ around its central energy, to a new direction due to diffraction by a crystal. This occurs when the crystal is oriented to fulfill the Bragg or Laue condition~\cite{voegeli2023, villanueva2023megahertz}. The Bragg or Laue condition is fulfilled when the incident X-ray beam intersects a family of lattice planes at a specific angle of incidence $\theta_\textrm{\scriptsize B}$ with respect to the beam direction and is dependent on the crystal material, the lattice plane, and the beam energy. The redirected portion of the beam (beamlet) travels at an exit angle of $2\theta_\textrm{\scriptsize B}$. On the other hand, amplitude splitting refers to the positioning of the crystal so that this condition is only fulfilled for a portion of the beam's footprint, i.e., only part of the crystal is in the beam path, allowing the rest of the beam to propagate unhindered~\cite{roling2012amplitude}. This strategy avoids absorption losses, facilitating higher acquisition rates, albeit at the cost of decreasing the \ac{FOV}. The feasibility of performing fast 3D imaging experiments with these splitting schemes is influenced by several factors, most importantly: stable and homogeneous illumination, which is essential for achieving sufficient image quality, photon flux, which determines signal-to-noise ratio and temporal resolution, and angular spacing of the viewpoints, which affects the quality of the volumetric reconstruction. These aspects will be elaborated on in the following, beginning with an emphasis on the quality of the crystals used to generate the beamlets.


Stable and homogeneous beamlet illumination is inherently dependent on the quality of the beam splitters. Perfect, dislocation-free crystals have been shown to offer the best performance and were therefore used for the fabrication of the beam splitters employed in this work. 
Additionally, the fabrication protocols can influence the choice of materials, as well as the specifications for their size and thickness. While the detailed fabrication process is beyond the scope of this work, we emphasize that it plays a critical role in the performance of the XMPI setup.
Even when the fabrication of the crystal is deemed successful, i.e., the crystals are mostly dislocation-free, radiation damage on the beam splitters during their exposure to X-rays can significantly degrade the image quality. To maintain stable and homogeneous illumination for high-quality imaging, thermal effects have to be considered. These effects can manifest as microscopic and macroscopic morphological changes in the crystal. Therefore, materials with high thermal conductivity and high melting point are preferred, as they are more resistant to heat-induced deformation and radiation damage, ensuring long-term stability and performance of the XMPI setup.

The photon flux carried by the beamlets is influenced mainly by the beam and crystal properties. 
Regarding the former, to increase the flux carried by the beamlets, it is desirable to have a beam with small divergence, due to the finite angular acceptance (Darwin width) of the crystal, and high spectral flux density, as typically illustrated in DuMond diagrams~\cite{Authier:book}.
Regarding the crystal properties, one has to optimize the choice of crystal characteristics such as the material's atomic number ($Z$), diffraction order, and the layout-introduced effects, i.e., flux losses from transmission through elements placed upstream of each beam splitter.
The atomic number of the crystal material affects the diffraction intensity - and thus the beamlet flux - since materials with higher $Z$ contain more electrons, increasing the probability of X-ray diffraction within the crystal’s extinction length \cite{Authier:book}. This flux directly impacts the utilized ADC range of the detector: higher diffraction efficiency, i.e., a larger amount of diffracted photons with respect to the incident direct beam, allows for higher acquisition rates while maintaining full detector performance.
However, while high-$Z$ materials offer stronger diffraction intensities, they also exhibit short extinction lengths and greater beam attenuation. This is particularly important because some of the beam splitters are arranged sequentially along the beam path, as will be discussed later.
These competing properties necessitate the use of thin implementations of high-$Z$ materials to balance diffraction efficiency and transmission.
In addition to material properties, diffraction efficiency is influenced by the Darwin width and the diffraction order of the crystal.
The former is an individual property of crystal materials, and a larger Darwin width will lead to diffracting a larger bandwidth of the spectrum. Higher diffraction orders enable exploring larger angles, which is desirable for maximizing the angular spacing between beamlets. 
However, this comes at the cost of reduced diffraction efficiency, especially due to the horizontal polarization of synchrotron and XFEL sources, which leads to significantly lower efficiency at Bragg angles near $45^\circ$ \cite{Authier:book}.

To address these constraints, we currently employ four beam splitters in our layout, as illustrated in Figure~\ref{fig:XMPI_setup}: three beam splitters are placed sequentially along the primary beam path, and a fourth beam splitter is placed off-axis with respect to the primary beam to redirect the beamlet generated by the first beam splitter towards the sample position, thereby increasing the angular spacing between the beamlets. All beam splitters are oriented to fulfill the Bragg condition, except for the second sequential beam splitter, which is oriented to fulfill the Laue condition.

We were taking into account the previously discussed aspects when choosing which crystals and diffraction planes to utilize in our \ac{XMPI} setup. As materials, we selected Ge and Si crystals, as these are among the most accessible materials that possess high purity and exhibit minimal dislocations. To generate the projections, all beam splitters employed spectral splitting, the crucial concept of \ac{XMPI}. Additionally, the first crystal was also an amplitude splitter to reduce absorption losses. We considered the Darwin widths of different diffraction planes to obtain sufficient flux on each projection, while at the same time being able to position the beam splitters accurately with the precision accessible with the nanopositioners they are mounted on. The details on the selected crystals for the different projections together with their corresponding Darwin widths are shown in Table~\ref{tab:XMPI_params}.

The precise positioning of the beam splitters is crucial to successfully perform \ac{XMPI} experiments. Each crystal must simultaneously satisfy its Bragg condition and be accurately aligned so that all beamlets intersect at a common point in space. Therefore, we employ a stack of six nanopositioners for each beam splitter, providing us with six degrees of freedom (three translations, one rotation, two goniometers). The translational stages provide a 1~nm closed-loop positioning resolution with the horizontal (vertical) stage covering a total travel range of 30~mm (8~mm). The rotation stage is capable of continuous rotation with $0.01$~m$^\circ$ (0.036~arcsec) resolution. The goniometers are able to move $\pm5^\circ$ with 1~\textmu$^\circ$ (0.0036~arcsec) resolution. All positioners are operated in closed-loop mode via manufacturer-provided \acp{API}, which are integrated into a custom Python control script.




\begin{table}[ht]

\caption{Experimental parameters of the two \ac{XMPI} experiments performed at ForMAX, MAX~IV.}
\label{tab:XMPI_params}
\begin{center}
    
\begin{tabular}{ l|l|l }
 Projection & Parameters & Values \\
 \hline 
 &energy & 16.5~keV\\
 &angular coverage & $48^\circ$ \\
 \hline
 \#1 &beam splitter & Si + Ge\\
 &diffraction mode & Bragg\\
 &out-of-plane  & 111 + 100\\
 &diffraction plane  & 111 + 400\\
 &Darwin width & 3.2~arcsec \& 3.0~arcsec\\
 &angle w.r.t. beam axis  & $-17.0^\circ$\\
 \hline
 \#2 &beam splitter & Si\\
 &diffraction mode & Laue\\
 &out-of-plane  & 111\\
 &diffraction plane  & 111\\
 &Darwin width & 3.2~arcsec\\
 &angle w.r.t. beam axis  & $13.7^\circ$\\
 \hline
 \#3 &beam splitter & Ge \\
 &diffraction mode & Bragg\\
 &out-of-plane  & 100 \\
 &diffraction plane  & 400 \\
 &Darwin width & 3.0~arcsec\\
 &angle w.r.t. beam axis  & $30.7^\circ$

\end{tabular}
\end{center}
\end{table}

\subsubsection{Detectors}
\label{sec:detectors}
The detection of each beamlet after the sample is performed using indirect detectors - detectors that first convert the X-rays to visible light, magnify the visible light through an optical microscope, and then record the images with a visible-light sensor. 
The detector setup is modular, allowing for different configurations of scintillators, magnification optics, and cameras, to accommodate a wide range of experimental requirements in terms of acquisition rate, spatial resolution, and \ac{FOV} of interests. The detectors' positions can be adjusted for each experiment as needed. They are positioned at the angle of the corresponding projection, i.e., $2\cdot\theta_\textrm{\scriptsize B}$ with respect to the beam axis, as listed in Table~\ref{tab:XMPI_params}. The distance between the sample and the detector may vary between projections, as it is primarily determined by the available physical space, typically ranging from 350~mm to 450~mm. 

The scintillator is mounted in a dedicated unit, designed to accommodate scintillators of $8~\textrm{mm}$ x $8~\textrm{mm}$ in size and up to 300~\textmu m thick. 
In this work, we use a 250~\textmu m thick  GaGG+ scintillator, known for its excellent high-yield ($45000~\textrm{ph/MeV}$), chosen to match the depth of focus for our desired spatial resolution of $\sim$8~\textmu m.
The microscopes are equipped with an objective holder featuring a motorized focusing unit, compatible with 5X to 20X objectives covering a broad wavelength spectrum, providing the flexibility to employ various cameras with different sensor types and scintillators. A high-resolution 5X objective is typically used. 
The focus unit is driven by a two-phase stepper motor, offering a $10~\textrm{mm}$ travel range with 1~\textmu m resolution.
Each microscope includes a tube lens holder with a motorized camera rotation unit. 
The nominal design is for a 1X tube lens.
The rotation unit covers a range of $90~\textrm{mrad}$ with an accuracy of $0.02~\textrm{mrad}$, and supports C- and F-mount cameras.

Additionally, the microscopes are mounted on a motorized positioning system for linear motion perpendicular to the beam-incidence axis and along the vertical axis, with travel ranges of $40~\textrm{mm}$ and $26~\textrm{mm}$, respectively, and a resolution of 5~\textmu m.

The choice of camera is motivated by the spatial and temporal resolution needs for the given experiment.
Commercial cameras often necessitate a trade-off between spatial and temporal resolution~\cite{Olbinado:17}.
According to the Nyquist-Shannon theorem~\cite{thapa2015nyquist}, a feature can be resolved when the sampling rate is at least twice the highest spatial frequency - corresponding to two pixels per feature.
Thus, higher spatial resolution requires smaller pixel sizes, which in turn have a lower signal-to-noise ratio and demand higher flux or longer exposure times, effectively limiting the achievable temporal resolutions.
Conversely, high temporal resolution requires high pixel sensitivity and fast readout schemes, with the latter often achieved by decreasing the effective sensor area at higher acquisition rates.

In this work, we deployed the \ac{XMPI} microscopes with two camera systems, the Photron Nova S16 for high temporal resolution, and the Andor Zyla 5.5 for high spatial resolution. 
Their specifications are summarized in Table~\ref{tab:cameras}. 
The Photron Nova S16 supports full-frame acquisitions ($1024~\times~1024$~pixels) at up to 16~kHz, and can reach $1.1$~MHz at a reduced frame ($128~\times~16$~pixels). 
The Andor Zyla 5.5 can reach 49 Hz with its full sensor ($2560~\times~2160$~pixels).

\begin{table}[ht]
\caption{Specifications of the cameras deployed for the \ac{XMPI} experiments discussed in the present paper.}
\label{tab:cameras}
\begin{center}
\begin{tabular}{llll}      
Camera    & Sensor Size (W x H)        & Pixel Size (W x H)       & ADC     \\
\hline
    Photron Nova S16    & 1024 pixels $\times$ 1024 pixels & 20~\textmu m $\times$ 20~\textmu m   & 12~bit\\
    Andor Zyla 5.5      & 2560 pixels $\times$ 2160 pixels & 6.5~\textmu m $\times$ 6.5~\textmu m & 16~bit\\
\end{tabular}
\end{center}
\end{table}

We have performed a detailed analysis of the spatial resolution of the optical system using the Photron Nova S16 at $40~\textrm{kHz}$, with the 250~\textmu m thick GaGG+ scintillator and 5X objective at an energy of 16.3~keV. We concluded that the Nyquist-Shannon theorem is the limiting factor, providing a spatial resolution of 8~\textmu m ($2\times 4$~\textmu m) \cite{yao2024DLSR}. The same considerations would mean a spatial resolution of 2.6~\textmu m ($2 \times 1.3$~\textmu m) for the Andor Zyla 5.5.

XMPI provides simultaneous illumination of the sample, and, therefore, a synchronous camera acquisition scheme is required.
A hardware trigger signal initiates the acquisition of each camera. 
This trigger can be configured to be issued i) from the sample if events of interest can be used as a trigger, e.g., tensile load value exceeding a minimum, or ii) from an external input like a hardware button trigger. 
The Zyla cameras are fully integrated in the ForMAX signal control system, ensuring synchronized hardware triggering using a PandABox~\cite{zhang17}.
The Photron cameras were operated in a chain configuration; one camera received the trigger signal and acted as the primary transmitter, triggering the acquisition of the two other cameras, which acted as receivers.

In this work, we refer to the Photron Nova S16 configuration as the ``high temporal resolution'' setup, and the Andor Zyla 5.5 configuration as the ``high spatial resolution'' setup.

\subsubsection{Sample}
\ac{XMPI} is compatible with a plethora of scientific samples and sample environments.
The X-ray energy used in an experiment is determined by the contrast and flux requirements of the sample under investigation, while the physical footprint of the sample environment dictates the necessary spacing between the beam splitters.
To ensure accurate alignment, the region of interest (ROI) within the sample is positioned at the beamlet intersection point using a stack of three linear motors - one for vertical and two for horizontal adjustments.

\subsection{Data processing and reconstruction} 
To reconstruct the 3D volume from the three recorded projections, we start by pre-processing the data. We perform conventional flat-field correction on the projections to reduce the inhomogeneities caused by the beam itself. For this, we record a series of flats, i.e., images without the sample in the beam, and average them. Then, we divide the projections by the average flat field~\cite{van2015FFC}.

For the actual reconstruction, we face the challenge of having a sparse number of views available. In conventional tomography, the Crowther criterion is used to determine the number of required projections $N_\theta$ based on the number of horizontal pixels $N_x$~\cite{jacobsen2019microscopy}:
\begin{equation}
    N_\theta = \frac{\pi}{2}N_x.
    \label{eq:crowther}
\end{equation}
In case of 1024~pixels, this corresponds to $\sim$1600 projections instead of the three projections offered with the presented layout of \ac{XMPI}. Consequently, traditional reconstruction algorithms such as filtered back projection (FBP) are unable to accurately reconstruct the volume. To perform 4D reconstructions (3D + time) from a sparse number of projections, one can use the recently developed self-supervised deep-learning tools 4D-ONIX and X-Hexplane that include the physics of X-ray interaction with matter.

Both algorithms directly reconstruct the 4D volume. 4D-ONIX operates by (i) integrating X-ray propagation physics into the model, (ii) employing a continuous representation of the sample, which describes the refractive index as a function of spatial and temporal coordinates, (iii) learning the latent features of the sample by generalizing across different experiments of similar sample dynamics, and (iv) using adversarial learning to enforce consistency between real and predicted projections~\cite{zhang2023onix, zhang20244donix,yao2025PhysONIX}. X-Hexplane operates in a similar manner but represents 4D as a factorized combination of orthogonal spatial and spatiotemporal feature planes, i.e., three spatial (XY, YZ, ZX) and three spatiotemporal (XT, YT, ZT)~\cite{cao2023hexplane}. By employing this tensorial representation, X-Hexplane improves memory efficiency and computational performance compared to 4D-ONIX. The detailed working principles of both algorithms are beyond the scope of this work, and the interested reader is referred to \cite{zhang20244donix} and \cite{hu2025super}. 

\begin{figure}[ht]
    \centering
    \includegraphics[scale=0.3]{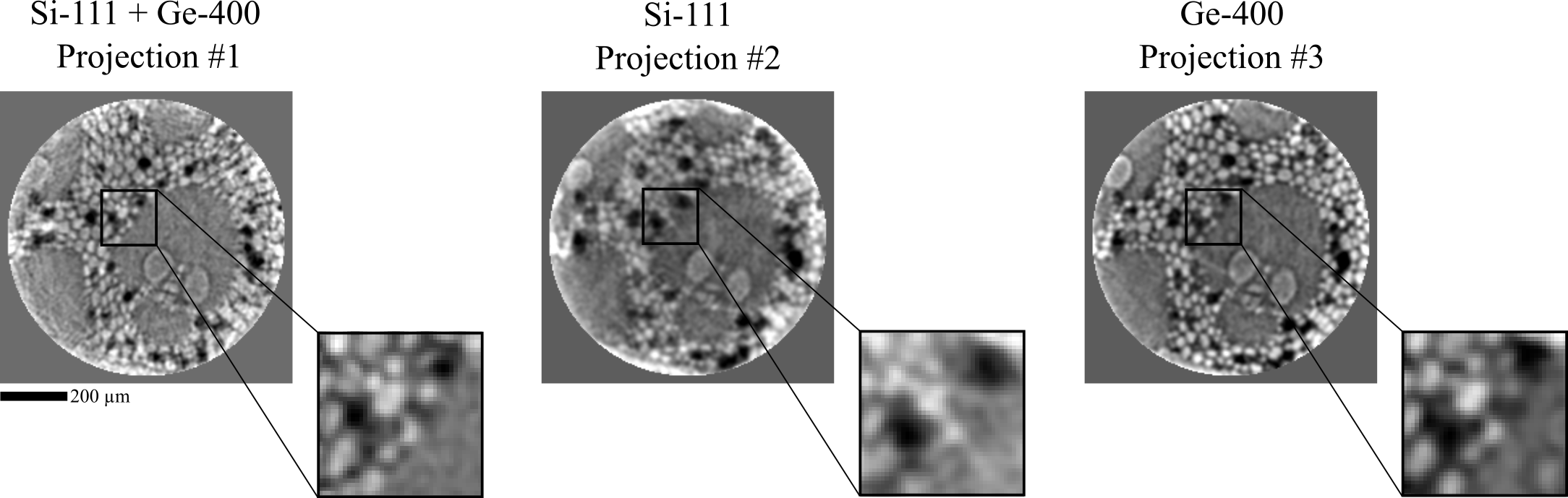}
    \caption{Reconstructed tomography slice of a bamboo rod recorded with each beamlet individually. The dataset was acquired with an acquisition speed of 6~kHz and a rotation speed of 72~deg/s. For the reconstruction, 349 projections over 180$^\circ$ were used follwing the Crowther criterion.}
    \label{fig:tomo}
\end{figure}

As of now, 4D-ONIX and X-Hexplane have been successfully applied to reconstruct simple dynamics, such as droplet collisions, in 4D. For a reconstruction of more complex sample systems, additional physics knowledge, such as support constraints or equations describing the underlying physical processes, can be highly beneficial. Having more prior information available, e.g., from several experiments, and/or a larger number of projections, also improves the reconstruction~\cite{zhang20244donix, yao2025PhysONIX}. Alternatively, we are working on enriching the spatial information by combining a sufficiently slow rotation with XMPI, if the sample process in question allows for it, thereby helping with the reconstruction~\cite{hu2025super}. We can also perform standard tomography to constrain either initial or final sample states. In Figure~\ref{fig:tomo} we demonstrate the quality of the beamlets by performing tomography of a bamboo rod with each of the three beamlets (Si-111 \& Ge-400, Si-111 and Ge-400) with an acquisition rate of 6~kHz and rotation speed of 72~deg/s, when using 349~projections over 180$^\circ$ for the reconstruction, following the Crowther criterion. It should be noted that the tomogram acquired with the Si-111 appears more blurry due to Laue diffraction accompanied by the Borrmann triangle effect, as will be discussed in Section~\ref{sec:Discussion}.

Depending on the goal of the experiment, a 4D reconstruction might not be necessary. Another example of employing \ac{XMPI} is particle tracking, where two projections are sufficient in order to identify individual particles and study their behavior~\cite{rosen2024xmpi}.

\section{Examples of scientific applications enabled by XMPI}
\label{sec:ScientificApplications}
Scientific applications with i) sensitive dynamics, such as fluidic samples and wood fibers, and ii) complex sample environments, are two cases of the previously mentioned studies that cannot be fully explored with in-situ and operando time-resolved tomography.
The potential of \ac{XMPI} lies in probing unexplored phenomena in such dynamics, and in this work, we are employing \ac{XMPI} for the study of (i) loading properties of wood fibers under different stages of Kraft cooking and (ii) particle suspension in multi-phase flows.

\subsection{Samples}
\subsubsection{Wood fibers' loading properties}
\label{subsec:wood_fibers_loadingProperties}
With increasing interest in replacing plastic-based materials, wood fibers are gaining importance in construction, pulp and paper, and advanced technology. 
To advance wood-based materials towards such applications, it is crucial to understand their mechanical behavior under external stress. 
\ac{XMPI} offers an opportunity to study in 3D the deformation process and breakage under load at spatiotemporal resolutions not possible with state-of-the-art methods. 
Moreover, it circumvents the misinterpretation of fiber mechanical behavior by using one projection. 

Here, we focused on loading properties of spruce wood fibers under different stages of Kraft cooking \cite{dang2016kraft}, a process typically used in industries to extract cellulose fibers. The sample size was roughly 500~\textmu m both in width and thickness, while the length was adjusted to 1~cm. Considering that the average size of a spruce fiber is between 30 and 50~\textmu m in diameter, the dynamics of more than 10 by 10 fibers can be observed simultaneously, where the deformation of wood fiber walls (around 5~\textmu m thick) and crack propagation within them would be detected. We needed to prioritize high temporal resolution during the measurements, since the events of interest were expected to occur on timescales shorter than $1~\textrm{ms}$. To this end, the high temporal resolution configuration of XMPI was deployed for $12.5~\textrm{kHz}$ acquisitions, at an energy of 16.5~keV.

\subsubsection{Multiphase flows}
\label{subsec:multiphaseFlows}
Understanding the behavior of multiphase flows is essential due to their prevalence in both natural and industrial processes, motivating the need to acquire volumetric 4D information (3D + time) without disruptive forces. For this reason, we performed \ac{XMPI} experiments to study multi-phase flow properties via particle tracking, only requiring two projections. For this purpose, we chose Projection~\#1 and \#3 to maximize the angular coverage.  We examined the motion of silver-coated hollow glass spheres (SHGS) with a diameter of 10~\textmu m through a glycerol solution, thereby minimizing gravity-induced effects on the flow. To accurately monitor the particles' flow on the micrometer scale, we prioritized achieving a high spatial resolution by deploying the high spatial resolution XMPI configuration while compromising at lower temporal resolutions with acquisitions of $40~\textrm{Hz}$, exploring flow rates of 0.1~mL/h, 0.2~mL/h, and 0.5~mL/h. To optimize the contrast within the flow system, we chose an energy of 16.5~keV~\cite{rosen2024xmpi}.

Such volumetric and time-resolved X-ray studies represent a critical advancement, paving the way for future experiments that can explore previously inaccessible research questions regarding, e.g., blood flow properties and food and material processing.

\section{Discussion} 
\label{sec:Discussion}
We deployed XMPI at ForMAX for two experimental studies. The studies were performed at 16.5~keV with two configuration setups, allowing us to explore distinct spatiotemporal regimes. The experimental setup, along with indicative data, is presented in Section~\ref{sec:exp}, followed by the discussion of current challenges and limitations in Section~\ref{sec:challenges}.

\subsection{Current Possibilities of XMPI at ForMAX}
\label{sec:exp}
Our experiments demonstrated the potential of \ac{XMPI} experiments at ForMAX on studies with high temporal (Experiment~\#1) and high spatial (Experiment~\#2) resolution requirements, utilizing the detector's full or near-full ADC range. The utilized ADC range serves as an indicator of photon statistics of our detector, helping us to optimize illumination by adjusting the acquisition rate and/or applying filtering. 

\subsubsection{Experiment \#1}
The high temporal resolution configuration of XMPI at ForMAX was deployed with the Photron Nova S16 cameras (12-bit) for the imaging of wood fiber breakage, which occurs on sub-millisecond timescales and requires above kHz acquisitions for capturing the event of interest.
The configuration offers a spatial resolution of approximately $8$~\textmu m on each projection (Section~\ref{sec:detectors}).
We deployed XMPI with three projections with the use of four beam splitters (Table~\ref{tab:XMPI_params}).

\begin{figure}[ht]
    \centering
    \includegraphics[scale=0.29]{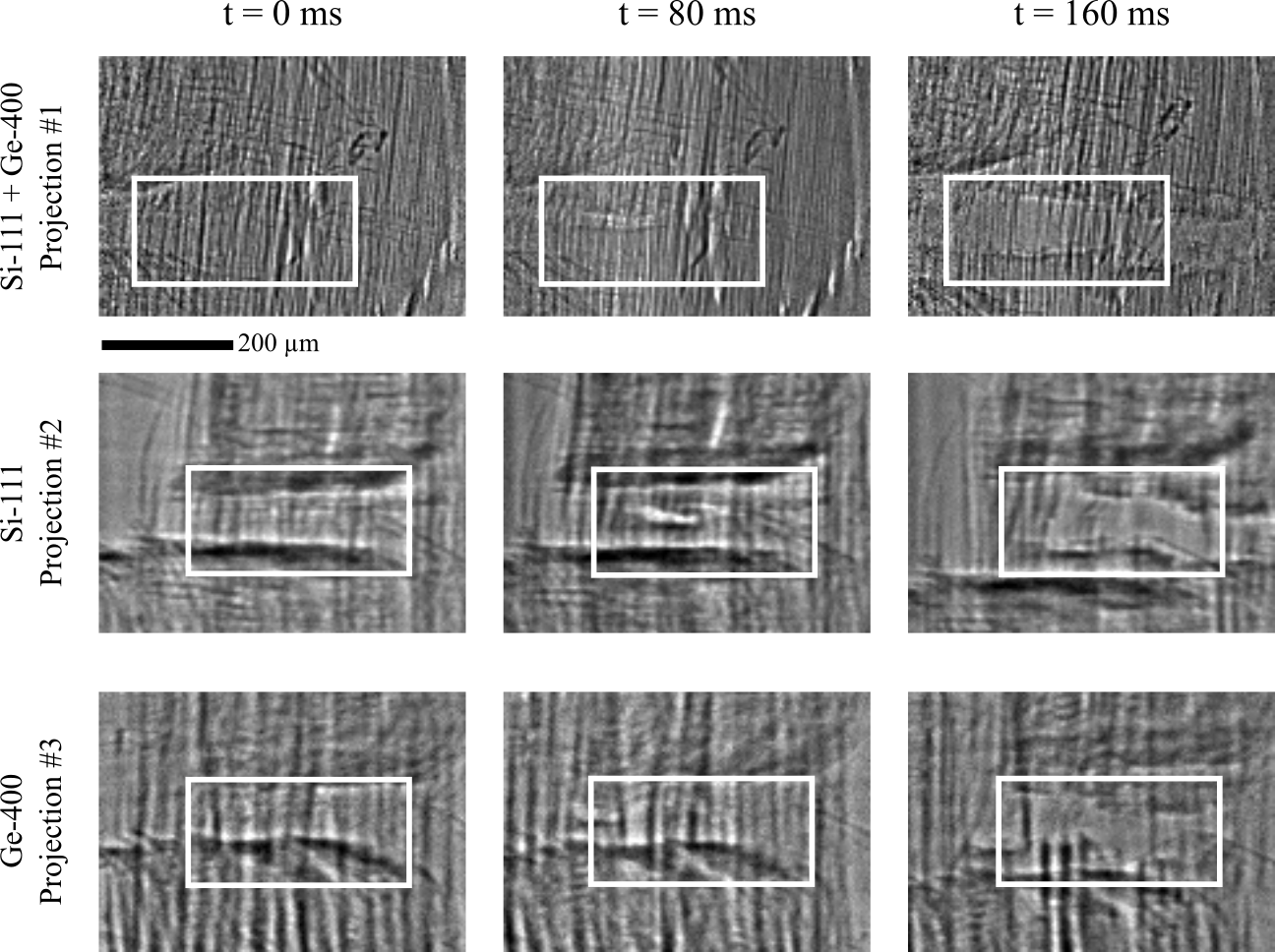}
    \caption{\ac{XMPI} projections of wood fiber failure recorded at 12.5~kHz with an energy of 16.5~keV. The failure region is highlighted by a box. Three representative time stamps were selected (0~ms, 80~ms and 160~ms). The projections were flat field corrected using conventional flat field correction. Projection~\#1 with $-17.0^\circ$, \#2 with $13.7^\circ$ and \#3 with $30.7^\circ$ with respect to the primary beam axis were generated with a recombiner (Si-111 + Ge-400), Si-111 and Ge-400, respectively.}
    \label{fig:fiber}
\end{figure}

Breakage of wood fibers is a dynamic process that requires this level of spatiotemporal resolution, as discussed in Section~\ref{subsec:wood_fibers_loadingProperties} and was thus deemed ideal for the demonstration of the high temporal capabilities of XMPI.
The evolution of the process was recorded successfully from three angularly spaced viewpoints over several seconds, with a temporal resolution of 160~\textmu s assuming Nyquist resolution.
Figure~\ref{fig:fiber} presents a selection of frames acquired during the experiment. The utilized ADC of the recorded data ranged from 10 to 12 bits. 

\newpage

\subsubsection{Experiment \#2}
\begin{figure}[ht]
    \centering
    \includegraphics[scale=0.29]{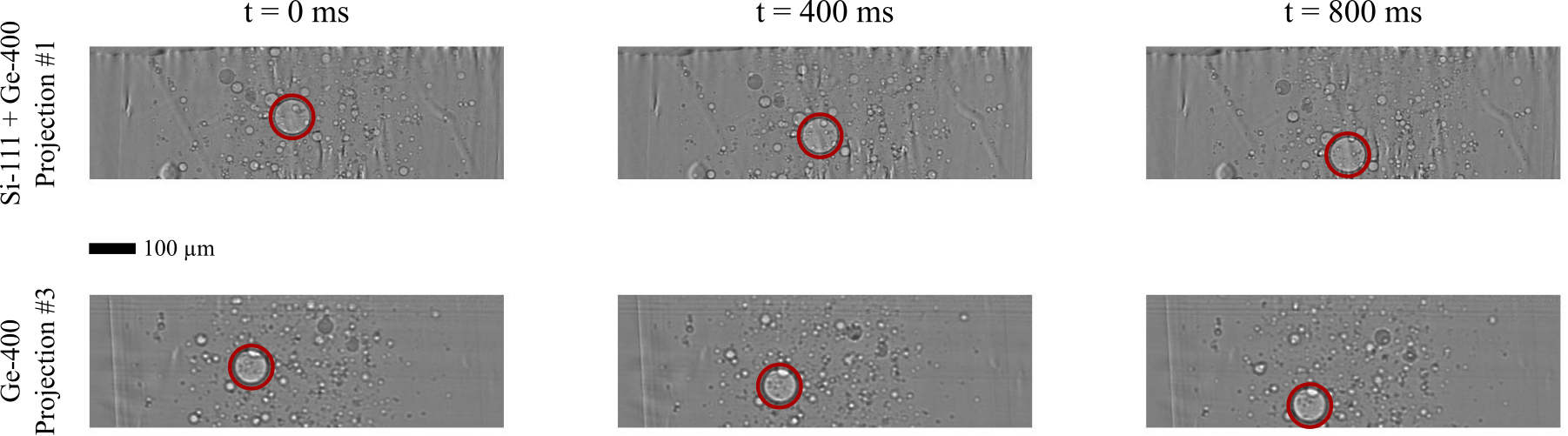}
    \caption{\ac{XMPI} projections of SHGS suspended in glycerol recorded at 40~Hz with an energy of 16.5~keV. Three representative time stamps were selected (0~ms, 400~ms and 800~ms). The projections were flat field corrected using conventional flat field correction. Projection~\#1 with $-17.0^\circ$ and \#3 with $30.7^\circ$ with respect to the primary beam axis were generated with a recombiner (Si-111 + Ge-400) and Ge-400, respectively. The same particle is highlighted in red in each frame to illustrate its movement with time.}
    \label{fig:fluid}
\end{figure}

The high spatial resolution configuration of XMPI at ForMAX was deployed with the Andor Zyla 5.5 cameras (16-bit) for the imaging of multi-phase flows. The configuration enables the resolution of features down to 2.6~\textmu m, meeting the requirements for this sample study. Images were recorded at an acquisition rate of 40~Hz with Projection~\#1 and \#3, as listed in Table~\ref{tab:XMPI_params}.
Figure~\ref{fig:fluid} shows selected frames from the two projections, with SHGS particles suspended in glycerol as the sample.
The full ADC range of the detector system was utilized in this configuration. The detailed analysis of the flow properties is out of the scope of this work but is discussed by Ros{\'e}n et al. \cite{rosen2024xmpi}.

\newpage
\subsection{Splitting Limitations}
\label{sec:challenges}
We are continuously striving to improve the \ac{XMPI} setup; further developing and optimizing the instrumentation. 
Improving the hardware will lead to higher-quality data and will enable better analysis methods, most importantly volume reconstructions, where it is needed. 

\begin{figure}[ht]
    \centering
    \includegraphics[scale=0.2]{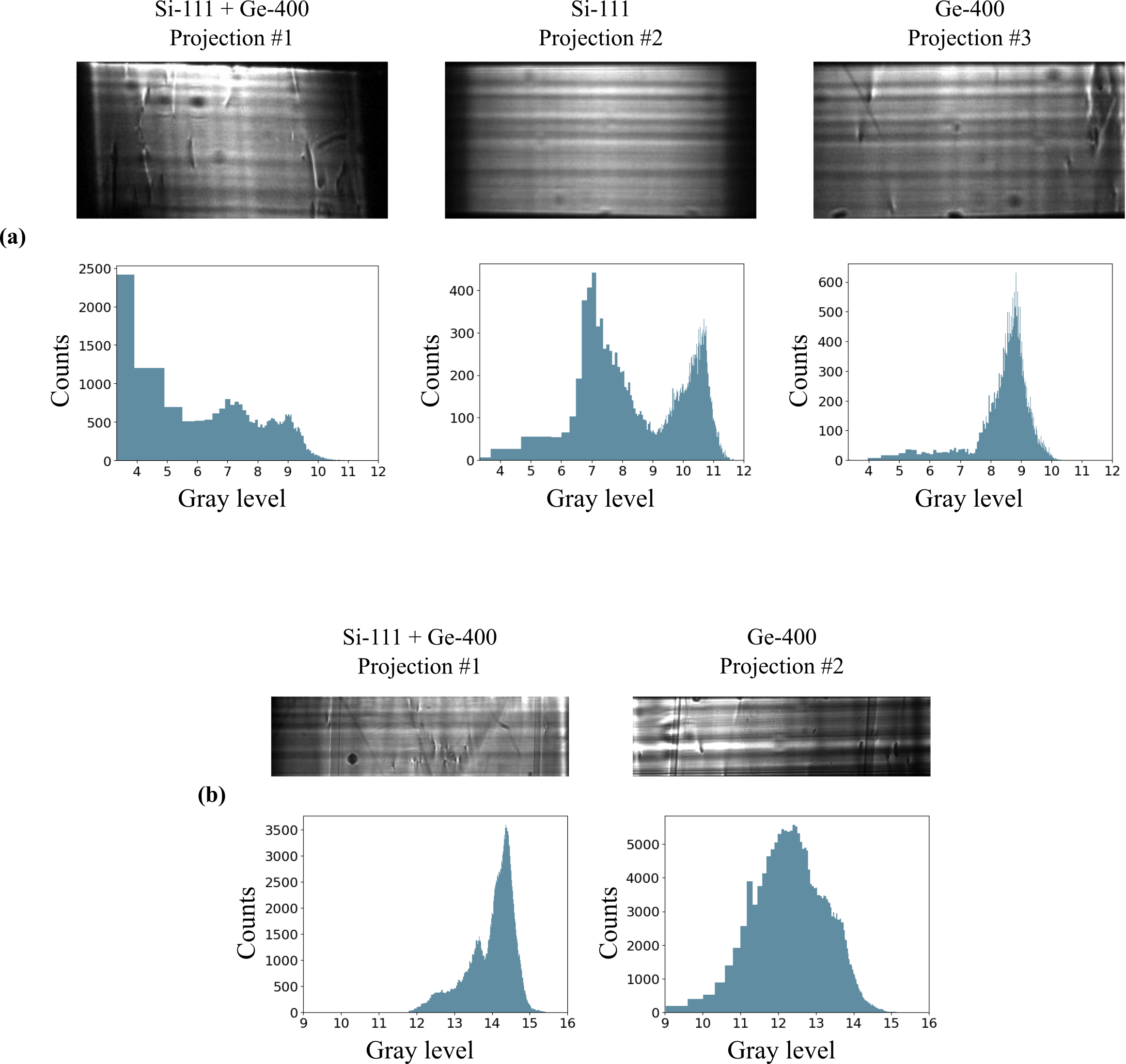}
    \caption{Flat field images of two different experiments acquired at 12.5~kHz (a) and 40~Hz (b) with corresponding histograms. The x-axis is log-scaled with base 2, and the tick labels indicate the exponent. (a) The high temporal resolution configuration covers an ADC range of 12~bit (Projection~\#1), 12~bit (Projection~\#2) and 11~bit (Projection~\#3). The full ADC range of the Photron Nova S16 is 12~bit. (b) The high spatiotemporal resolution configuration covers an ADC range of 16~bits for both projections, i.e., the full ADC range of the Andor Zyla 5.5.}
    \label{fig:histo}
\end{figure}

High-quality projections, homogeneous illumination with as few artifacts as possible, are required by XMPI. 
These strongly depend on the beam splitters. 
Such a homogeneous intensity can be disturbed because of diffraction at slightly different angles in the active area of the crystal, induced by imperfect crystals or non-optimal clamping of the crystals. In fact, the fabrication of perfect crystals that satisfy our requirements when it comes to dimensions and purity, as well as the absence of dislocations, is challenging. Difficulties with inhomogeneous illumination are depicted in Figure~\ref{fig:histo}(a) and (b), where flat-field images of the projections with their corresponding histograms are presented. Some parts of the flats are saturated due to uneven illumination, resulting in certain areas of the image appearing significantly brighter than others. An approach to overcome fabrication challenges is to employ Laue rather than Bragg diffraction. Using a beam splitter in the Bragg configuration requires a larger active area of the crystal, as the angle between the beam splitter and the direct incident beam is small, i.e., the beam has a large footprint, and more surface is needed to diffract the whole beam. On the contrary, beam splitters used in symmetric Laue geometry can be exposed to a smaller beam footprint, relaxing the requirements for the size of the active area. This is important because it is less demanding to manufacture smaller crystals with the same technical specifications, which reduces the risks of imperfections. 
However, it should be noted that Laue diffraction introduces some blurring in the image due to the Borrmann triangle effect~\cite{Authier:book, asimakopoulou2024xmpi}. 
Moreover, to avoid strains on the crystal surface, it is important to optimize clamping schemes and consider strain relief cuts~\cite{bellucci2024crystals}. 

Here, we used the beam splitter of Projection~\#2 in symmetric Laue geometry. We employed a silicon crystal formed as a thin membrane supported by an integral frame. The membrane was fabricated by adapting processes developed for crystals used to study high-energy particle–crystal interactions~\cite{mazzolari2013fabrication, scandale2014mirroring, germogli2015manufacturing}. A silicon-oxide hard mask was deposited on 100-mm-diameter 0.5-mm-thick silicon wafers through low-pressure chemical vapor deposition techniques and patterned photolithographically. The exposed silicon was then thinned in an isotropic wet etchant~\cite{madou2011fundamentals}. In our process, the attainable active area was limited to a few tens of mm$^2$, and the thickness uniformity across the aperture was $\sim$~5~\textmu m. Isotropic undercut and etch-bath hydrodynamics imposed practical limits on window size and thickness uniformity~\cite{kuiken1984etching}. After thinning, the hard mask was removed by wet etching, and the wafer was diced to release individual crystals. The frame enabled safe handling and, crucially, prevented clamp-induced stresses from propagating into the active diffracting area. To decouple the clamp from the active region, we machined strain-relief cuts in the frame~\cite{bellucci2024crystals}, forming a slotted compliant suspension that localized mechanical loads due to clamping and mitigated stress transfer to the membrane. 

The high photon flux required to perform \ac{XMPI} experiments with sufficiently intense diffracted X-ray beams will induce heat load on the beam splitters. Depending on the material and photon flux, this can appear as warping issues or vibrations, visible at short time scales, or, in extreme cases, as morphological changes of the crystal. Therefore, one might consider thermalization and/or cooling of the crystals when acquiring images.

Another limitation of \ac{XMPI} is the angular coverage of the setup. Obtaining higher diffraction angles requires higher diffraction orders, which comes with the cost of lower diffraction efficiency~\cite{Authier:book, Yashiro:2023, bellucci2024crystals}. In addition, \ac{XMPI} requires a volume reconstruction from a sparse number of projections, something that is possible with our algorithms for simple dynamics. However, more projections would help to improve the reconstructions further. In principle, the number of projections can still be increased \cite{voegeli2023, voegeli2024multibeam, Yahsiro10676953}, but at a certain point, we will reach a limit in terms of physical space available to fit our equipment, remaining bandwidth of the incident's beam undulator spectrum for spectral splitting (smaller Darwin widths will enable more projections at the cost of achievable spatiotemporal resolution per projection) and remaining beam size when employing amplitude splitting. Therefore, we are also integrating the possibility of an additional slow rotation of the sample (if the process allows it) in our \ac{XMPI} experiments, thereby increasing the number of acquired projections.
Finally, the performance of our algorithms in the volumetric time-resolved reconstruction of dynamic processes can be further increased by the 3D characterization (tomography) of the initial and final states of the studied samples, whenever possible.


\section{Conclusion and Outlook}
\label{sec:ConclusionAndOutlook}
In this work, we presented the commissioning and potential of our \ac{XMPI} setup at the ForMAX beamline at MAX~IV. \ac{XMPI} is a technique that can be used to retrieve 4D information on processes of interest, without the need for sample rotation, a feature that overcomes known limitations in present state-of-the-art methods such as tomography.
This novelty enables temporally resolved volumetric imaging of presently unexplored phenomena in rotation-sensitive samples, samples that require complicated study environments, samples with fast (above kHz) dynamics, or single-shot phenomena.
In \ac{XMPI}, several projections are acquired simultaneously by splitting the direct incident X-ray beam into beamlets that intersect at the sample position by employing beam splitters. The beam quality at ForMAX provided a unique environment for the successful commissioning of the technique due to its high spectral flux density.

We deployed XMPI in two different spatiotemporal regimes: a high temporal resolution configuration for at least 12.5 kHz acquisitions with the ability to distinguish 8~\textmu m features and a high spatial resolution configuration for distinguishing 2.6~\textmu m features with $40~\textrm{Hz}$ acquisitions, in the recorded projections. The commissioning of the two configurations was done during our reported work on Experiment \#1, for the study of wood fiber failure when exposed to external forces, and Experiment \#2, for the study of opaque multi-phase particle flows, where we performed in-flow particle tracking.

We discussed the presently identified main challenges of the technique, i.e., i) the fabrication of perfect crystals, ii) strains and artifacts on the crystals, and iii) a limited number of projections. 
We have identified the difficulty to address i), as we depend on having dislocation-free, perfect crystals that are large enough to capture the full X-ray footprint and have a thickness that considers the extinction length and still lets photons transmit towards the next beam splitter. Regarding (ii), we conclude that alternative clamping methods and strain relief cuts, as well as employing beam splitters in Laue geometry, will perform better. Furthermore, thermalization and cooling mechanisms should be considered to decrease the heat load on the crystals. To tackle (iii), we have the possibility to implement a slow rotation of the sample to acquire more projections, which will help with the volume reconstruction. In the future, we plan to test more materials as beam splitters, allowing us to explore different geometries and diffraction efficiencies. Moreover, we may consider more efficient signal detection schemes with a higher signal-to-noise ratio using direct conversion detectors, instead of scintillator-based microscopes, relaxing requirements when it comes to the flux of the beamlets without compromising on spatiotemporal resolution.

In conclusion, \ac{XMPI} presents exciting opportunities for 4D imaging, enabling the study of phenomena that remain unexplored due to the limitations of present state-of-the-art methods. The deployment of \ac{XMPI} at ForMAX allowed the exploration of its full potential at the beamline and we foresee that it opens the potential for applications that span a broad spectrum of research areas, including blood flow analysis, additive manufacturing, fracture behavior in composites, failure mechanisms in fibers and the compression dynamics of foams, to name just a few. This advancement opens up a new era for 4D imaging, paving the way for discoveries across diverse scientific research questions.

\begin{acknowledgements}
We are thankful to Saeed Davoodi for his significant contribution to the sample preparation and participation in Experiment~\#1. We are grateful to Zdenek Matej for his support and access to the computing clusters at MAX~IV. We acknowledge the MAX IV Laboratory for beamtime on the ForMAX beamline under proposals 20220975, 20231192, 20241593, and 20241696. Research conducted at MAX IV, a Swedish national user facility, is supported by Vetenskapsrådet (Swedish Research Council, VR) under contract 2018-07152, Vinnova (Swedish Governmental Agency for Innovation Systems) under contract 2018-04969 and Formas under contract 2019-02496. Portions of the text in this manuscript were revised with the assistance of Copilot and ChatGPT, a large language model developed by OpenAI, to improve clarity and readability.
\end{acknowledgements}

\begin{funding}
We are grateful for the funding and support from the European Research Council (ERC Starting Grant 3DX-FLASH, 948426).
\end{funding}

\ConflictsOfInterest{The authors declare no conflicts of interest.
}

\DataAvailability{The data supporting the results of this study are available within this article.}

\bibliography{iucr} 

@article{Mokso2015_StereoDualEnergy,
author = "Mokso, Rajmund and Oberta, Peter",
title = "{Simultaneous dual-energy X-ray stereo imaging}",
journal = "Journal of Synchrotron Radiation",
year = "2015",
volume = "22",
number = "4",
pages = "1078--1082",
month = "Jul",
doi = {10.1107/S1600577515006554}
}

@article{Oberta2013_Splitters,
title = {A Laue–Bragg monolithic beam splitter for efficient X-ray 2-beam imaging},
journal = {Nuclear Instruments and Methods in Physics Research Section A: Accelerators, Spectrometers, Detectors and Associated Equipment},
volume = {703},
pages = {59-63},
year = {2013},
issn = {0168-9002},
doi = {https://doi.org/10.1016/j.nima.2012.11.042},
author = {P. Oberta and R. Mokso}
}

@article{eriksson2014DLSR,
  title={Diffraction-limited storage rings--a window to the science of tomorrow},
  author={Eriksson, Mikael and Van der Veen, J Friso and Quitmann, Christoph},
  journal={Synchrotron Radiation},
  volume={21},
  number={5},
  pages={837--842},
  year={2014},
  publisher={International Union of Crystallography}
}

@article{raimondi2023DLSR,
  title={Toward a diffraction limited light source},
  author={Raimondi, Pantaleo and Liuzzo, Simone Maria},
  journal={Physical Review Accelerators and Beams},
  volume={26},
  number={2},
  pages={021601},
  year={2023},
  publisher={APS}
}

@article{villanova2017fast,
  title={Fast in situ 3D nanoimaging: a new tool for dynamic characterization in materials science},
  author={Villanova, Julie and Daudin, R{\'e}mi and Lhuissier, Pierre and Jauffres, David and Lou, Siyu and Martin, Christophe L and Labour{\'e}, Sylvain and Tucoulou, R{\'e}mi and Mart{\'\i}nez-Criado, Gema and Salvo, Luc},
  journal={Materials Today},
  volume={20},
  number={7},
  pages={354--359},
  year={2017},
  publisher={Elsevier}
}

@article{XrayTomographyTomoscopyReview,
author = {García-Moreno, Francisco and Neu, Tillmann Robert and Kamm, Paul Hans and Banhart, John},
title = {X-ray Tomography and Tomoscopy on Metals: A Review},
journal = {Advanced Engineering Materials},
volume = {25},
number = {8},
pages = {2201355},
year = {2023}
}

@article{yao2024DLSR,
  title={New opportunities for time-resolved imaging using diffraction-limited storage rings},
  author={Yao, Zisheng and Rogalinski, Julia and Asimakopoulou, Eleni Myrto and Zhang, Yuhe and Gordeyeva, Korneliya and Atoufi, Zhaleh and Dierks, Hanna and McDonald, Samuel and Hall, Stephen and Wallentin, Jesper and others},
  journal={Synchrotron Radiation},
  volume={31},
  number={5},
  pages={1299--1307},
  year={2024},
  publisher={International Union of Crystallography}
}

@article{Maire2014,
  author = {E. Maire and P. J. Withers},
  title ={Quantitative X-ray tomography},
  journal = {International Materials Reviews},
  volume = {59},
  number = {1},
  pages = {1-43},
  year = {2014},
  doi = {10.1179/1743280413Y.0000000023},
}

@article{nygaard2024formax,
  title={ForMAX--a beamline for multiscale and multimodal structural characterization of hierarchical materials},
  author={Nyg{\aa}rd, K and McDonald, SA and Gonz{\'a}lez, JB and Haghighat, VAHID and Appel, Christian and Larsson, EMANUEL and Ghanbari, REZA and Viljanen, MIRA and Silva, Jos{\'e} and Malki, SULEYMAN and others},
  journal={Synchrotron Radiation},
  volume={31},
  number={2},
  pages={363--377},
  year={2024},
  publisher={International Union of Crystallography}
}

@article{Moreno:2021,
author = {García-Moreno, Francisco and Kamm, Paul Hans and Neu, Tillmann Robert and Bülk, Felix and Noack, Mike Andreas and Wegener, Mareike and von der Eltz, Nadine and Schlepütz, Christian Matthias and Stampanoni, Marco and Banhart, John},
title = {Tomoscopy: Time-Resolved Tomography for Dynamic Processes in Materials},
journal = {Advanced Materials},
volume = {33},
number = {45},
pages = {2104659},
year = {2021}
}

@article{schropp2015imaging,
  title={Imaging shock waves in diamond with both high temporal and spatial resolution at an XFEL},
  author={Schropp, Andreas and Hoppe, Robert and Meier, Vivienne and Patommel, Jens and Seiboth, Frank and Ping, Yuan and Hicks, Damien G and Beckwith, Martha A and Collins, Gilbert W and Higginbotham, Andrew and others},
  journal={Scientific reports},
  volume={5},
  number={1},
  pages={11089},
  year={2015},
  publisher={Nature Publishing Group UK London}
}

@article{kumar2016strength,
  title={Strength improvement of glass substrates by using surface nanostructures},
  author={Kumar, Amarendra and Kashyap, Kunal and Hou, Max T and Yeh, J Andrew},
  journal={Nanoscale Research Letters},
  volume={11},
  pages={1--7},
  year={2016},
  publisher={Springer}
}

@article{Hoshino2013,
doi = {10.1088/1748-0221/8/05/C05002},
year = {2013},
month = {may},
publisher = {},
volume = {8},
number = {05},
pages = {C05002},
author = {M Hoshino and T Sera and K Uesugi and N Yagi},
title = {Development of X-ray triscopic imaging system towards  three-dimensional measurements of dynamical samples},
journal = {Journal of Instrumentation},
}

@article{Villanueva-Perez:18,
author = {Pablo Villanueva-Perez and B. Pedrini and R. Mokso and P. Vagovic and V. A. Guzenko and S. J. Leake and P. R. Willmott and P. Oberta and C. David and H. N. Chapman and M. Stampanoni},
journal = {Optica},
number = {12},
pages = {1521--1524},
publisher = {Optica Publishing Group},
title = {Hard x-ray multi-projection imaging for single-shot approaches},
volume = {5},
year = {2018}
}

@article{duarte2019computed,
  title={Computed stereo lensless X-ray imaging},
  author={Duarte, J and Cassin, R and Huijts, J and Iwan, B and Fortuna, F and Delbecq, L and Chapman, H and Fajardo, M and Kovacev, M and Boutu, W and others},
  journal={Nature photonics},
  volume={13},
  number={7},
  pages={449--453},
  year={2019},
  publisher={Nature Publishing Group UK London}
}

@article{Bellucci:23,
author = {Valerio Bellucci and Marie-Christine Zdora and Ladislav Mike\v{s} and \v{S}arlota Birn\v{s}teinov\'{a} and Peter Oberta and Marco Romagnoni and Andrea Mazzolari and Pablo Villanueva-Perez and Rajmund Mokso and Christian David and Mikako Makita and Silvia Cipiccia and Jozef Uli\v{c}n\'{y} and Alke Meents and Adrian P. Mancuso and Henry N. Chapman and Patrik Vagovi\v{c}},
journal = {Opt. Express},
number = {11},
pages = {18399--18406},
publisher = {Optica Publishing Group},
title = {Hard X-ray stereographic microscopy for single-shot differential phase imaging},
volume = {31},
month = {May},
year = {2023},
doi = {10.1364/OE.492137},
}

@article{voegeli2023,
  title={Multi-beam X-ray optical system for high-speed tomography using a $\sigma$-polarization diffraction geometry},
  author={Voegeli, Wolfgang and Liang, Xiaoyu and Shirasawa, Tetsuroh and Arakawa, Etsuo and Hyodo, Kazuyuki and Kudo, Hiroyuki and Yashiro, Wataru},
  journal={Applied Physics Express},
  volume={16},
  number={7},
  pages={072007},
  year={2023},
  publisher={IOP Publishing}
}

@article{voegeli2024multibeam,
  title={Multibeam X-ray tomography optical system for narrow-energy-bandwidth synchrotron radiation},
  author={Voegeli, Wolfgang and Takayama, Haruki and Liang, Xiaoyu and Shirasawa, Tetsuroh and Arakawa, Etsuo and Kudo, Hiroyuki and Yashiro, Wataru},
  journal={Applied Physics Express},
  volume={17},
  number={3},
  pages={032002},
  year={2024},
  publisher={IOP Publishing}
}

@INPROCEEDINGS{Yahsiro10676953,
  author={Yashiro, Wataru and Liang, Xiaoyu and Abukawa, Tadashi and Voegeli, Wolfgang and Arakawa, Etsuo and Shirasawa, Tetsuroh and Kajiwara, Kentaro and Kudo, Hiroyuki},
  booktitle={2024 Conference on Lasers and Electro-Optics Pacific Rim (CLEO-PR)}, 
  title={Proof-of-Concept of Millisecond-Order-Temporal-Resolution 4D X-ray Tomography with Multibeam X-ray Imaging System}, 
  year={2024},
  volume={},
  number={},
  pages={1-3},
  doi={10.1109/CLEO-PR60912.2024.10676953}
}

@INPROCEEDINGS{Sumiishi10676491,
  author={Sumiishi, Hiroki and Voegeli, Wolfgang and Kajiwara, Kentaro and Urushihara, Yoshimasa and Liang, Xiaoyu and Kudo, Hiroyuki and Yashiro, Wataru},
  booktitle={2024 Conference on Lasers and Electro-Optics Pacific Rim (CLEO-PR)}, 
  title={Multi-Projection Imaging of a Woodlouse with an Improved Multibeam X-Ray Optical System}, 
  year={2024},
  volume={},
  number={},
  pages={1-4},
  doi={10.1109/CLEO-PR60912.2024.10676491}
}

@article{asimakopoulou2024xmpi,
  title={Development towards high-resolution kHz-speed rotation-free volumetric imaging},
  author={Asimakopoulou, Eleni Myrto and Bellucci, Valerio and Birnsteinova, Sarlota and Yao, Zisheng and Zhang, Yuhe and Petrov, Ilia and Deiter, Carsten and Mazzolari, Andrea and Romagnoni, Marco and Korytar, Dusan and others},
  journal={Optics Express},
  volume={32},
  number={3},
  pages={4413--4426},
  year={2024},
  publisher={Optica Publishing Group}
}

@article{yao2025PhysONIX,
  title={Physics-informed 4D X-ray image reconstruction from ultra-sparse spatiotemporal data},
  author={Yao, Zisheng and Zhang, Yuhe and Hu, Zhe and Kl{\"o}fkorn, Robert and Ritschel, Tobias and Villanueva-Perez, Pablo},
  journal={Measurement Science and Technology},
  year={2025}
}

@article{villanueva2023megahertz,
  title={Megahertz x-ray multi-projection imaging},
  author={Villanueva-Perez, Pablo and Bellucci, Valerio and Zhang, Yuhe and Birnsteinova, Sarlota and Graceffa, Rita and Adriano, Luigi and Asimakopoulou, Eleni Myrto and Petrov, Ilia and Yao, Zisheng and Romagnoni, Marco and others},
  journal={arXiv preprint arXiv:2305.11920},
  year={2023}
}

@inproceedings{roling2012amplitude,
  title={Design of an X-ray split-and delay-unit for the European XFEL},
  author={Roling, Sebastian and Samoylova, Liubov and Siemer, Bj{\"o}rn and Sinn, Harald and Siewert, Frank and Wahlert, Frank and W{\"o}stmann, Michael and Zacharias, Helmut},
  booktitle={X-Ray Free-Electron Lasers: Beam Diagnostics, Beamline Instrumentation, and Applications},
  volume={8504},
  pages={34--43},
  year={2012},
  organization={SPIE}
}

@book{Authier:book,
  author = "André Authier",
  title  =  "Dynamical Theory of X-Ray Diffraction",
  publisher = {International Union of Crystallography},
  year   = 2001 
}

@article{Olbinado:17,
author = {Margie P. Olbinado and Xavier Just and Jean-Louis Gelet and Pierre Lhuissier and Mario Scheel and Patrik Vagovic and Tokushi Sato and Rita Graceffa and Joachim Schulz and Adrian Mancuso and John Morse and Alexander Rack},
journal = {Opt. Express},
number = {12},
pages = {13857--13871},
publisher = {Optica Publishing Group},
title = {MHz frame rate hard X-ray phase-contrast imaging using synchrotron radiation},
volume = {25},
month = {Jun},
year = {2017},
doi = {10.1364/OE.25.013857},
}

@article{thapa2015nyquist,
  title={Less is more: compressive sensing in optics and image science},
  author={Thapa, Damber and Raahemifar, Kaamran and Lakshminarayanan, Vasudevan},
  journal={Journal of Modern Optics},
  volume={62},
  number={6},
  pages={415--429},
  year={2015},
  publisher={Taylor \& Francis}
}

@book{jacobsen2019microscopy,
  title={X-ray Microscopy},
  author={Jacobsen, Chris},
  year={2019},
  publisher={Cambridge University Press}
}

@article{zhang2023onix,
  title={ONIX: An X-ray deep-learning tool for 3D reconstructions from sparse views},
  author={Zhang, Yuhe and Yao, Zisheng and Ritschel, Tobias and Villanueva-Perez, Pablo},
  journal={Applied Research},
  volume={2},
  number={4},
  pages={e202300016},
  year={2023},
  publisher={Wiley Online Library}
}

@article{zhang20244donix,
  title={4D-ONIX for reconstructing 3D movies from sparse X-ray projections via deep learning},
  author={Zhang, Yuhe and Yao, Zisheng and Kl{\"o}fkorn, Robert and Ritschel, Tobias and Villanueva-Perez, Pablo},
  journal={Communications Engineering},
  volume={4},
  number={1},
  pages={1--12},
  year={2025},
  publisher={Nature Publishing Group}
}

@inproceedings{cao2023hexplane,
  title={Hexplane: A fast representation for dynamic scenes},
  author={Cao, Ang and Johnson, Justin},
  booktitle={Proceedings of the IEEE/CVF Conference on Computer Vision and Pattern Recognition},
  pages={130--141},
  year={2023}
}

@article{hu2025super,
  title={Super Time-Resolved Tomography},
  author={Hu, Zhe and Yao, Zisheng and Josefsson, Kalle and Garc{\'\i}a-Moreno, Francisco and Makowska, Malgorzata and Zhang, Yuhe and Villanueva-Perez, Pablo},
  journal={Advanced Science},
  pages={e11933},
  year={2025},
  publisher={Wiley Online Library}
}

@article{rosen2024xmpi,
  title={Synchrotron X-Ray Multi-Projection Imaging for Multiphase Flow},
  author={Ros{\'e}n, Tomas and Yao, Zisheng and Tejbo, Jonas and Wegele, Patrick and Rogalinski, Julia K and Nilsson, Frida and Mom, Kannara and Hu, Zhe and McDonald, Samuel A and Nyg{\aa}rd, Kim and others},
  journal={arXiv preprint arXiv:2412.09368},
  year={2024}
}

@article{dang2016kraft,
  title={The impact of ionic strength on the molecular weight distribution (MWD) of lignin dissolved during softwood kraft cooking in a flow-through reactor},
  author={Dang, Binh TT and Brelid, Harald and Theliander, Hans},
  journal={Holzforschung},
  volume={70},
  number={6},
  pages={495--501},
  year={2016},
  publisher={De Gruyter}
}

@article{bellucci2024crystals,
  title={Development of crystal optics for X-ray multi-projection imaging for synchrotron and XFEL sources},
  author={Bellucci, Valerio and Birnsteinova, Sarlota and Sato, Tokushi and Letrun, Romain and Koliyadu, Jayanath C. P. and Kim, Chan and Giovanetti, Gabriele and Deiter, Carsten and Samoylova, Liubov and Petrov, Ilia and Lopez Morillo, Luis and Graceffa, Rita and Adriano, Luigi and Huelsen, Helge and Kollmann, Heiko and Tran Calliste, Thu Nhi and Korytar, Dusan and Zaprazny, Zdenko and Mazzolari, Andrea and Romagnoni, Marco and Asimakopoulou, Eleni Myrto and Yao, Zisheng and Zhang, Yuhe and Ulicny, Jozef and Meents, Alke and Chapman, Henry N. and Bean, Richard and Mancuso, Adrian and Villanueva-Perez, Pablo and Vagovic, Patrik},
  journal={Synchrotron Radiation},
  volume={31},
  number={6},
  pages={1534--1550},
  year={2024},
  publisher={International Union of Crystallography}
}

@article{Yashiro:2023,
doi = {10.35848/1882-0786/ace0f2},
year = {2023},
month = {jul},
publisher = {IOP Publishing},
volume = {16},
number = {7},
pages = {072001},
author = {Xiaoyu Liang and Wolfgang Voegeli and Hiroyuki Kudo and Etsuo Arakawa and Tetsuroh Shirasawa and Kentaro Kajiwara and Tadashi Abukawa and Wataru Yashiro},
title = {Sub-millisecond 4D X-ray tomography achieved with a multibeam X-ray imaging system},
journal = {Applied Physics Express}
}

@INPROCEEDINGS{zhang17,
   author = "S.~Zhang and Y.~M.~Abiven and J.~Bisou and G.~Renaud and G.~Thibaux and F.~Ta and S.~Minolli and F.~Langlois",
   year = "2017",
   booktitle = "Proceedings of the 16th International Conference on Accelerator and Large Experimental Physics Control Systems (ICALEPCS 2017)",
   pages = "143-150"
}

@article{van2015FFC,
  title={Dynamic intensity normalization using eigen flat fields in X-ray imaging},
  author={Van Nieuwenhove, Vincent and De Beenhouwer, Jan and De Carlo, Francesco and Mancini, Lucia and Marone, Federica and Sijbers, Jan},
  journal={Optics express},
  volume={23},
  number={21},
  pages={27975--27989},
  year={2015},
  publisher={Optical Society of America}
}

@article{scandale2014mirroring,
  title={Mirroring of 400 GeV/c protons by an ultra-thin straight crystal},
  author={Scandale, W and Arduini, G and Butcher, M and Cerutti, F and Gilardoni, S and Lechner, A and Losito, R and Masi, A and Metral, E and Mirarchi, D and others},
  journal={Physics Letters B},
  volume={734},
  pages={1--6},
  year={2014},
  publisher={Elsevier}
}

@article{mazzolari2013fabrication,
  title={Fabrication of large area silicon nanothickness membranes for channeling experiments},
  author={Mazzolari, Andrea and Guidi, Vincenzo and De Salvador, Davide and Bacci, Luca},
  journal={Nuclear Instruments and Methods in Physics Research Section B: Beam Interactions with Materials and Atoms},
  volume={309},
  pages={130--134},
  year={2013},
  publisher={Elsevier}
}

@article{germogli2015manufacturing,
  title={Manufacturing and characterization of bent silicon crystals for studies of coherent interactions with negatively charged particles beams},
  author={Germogli, Giacomo and Mazzolari, Andrea and Bandiera, Laura and Bagli, Enrico and Guidi, Vincenzo},
  journal={Nuclear Instruments and Methods in Physics Research Section B: Beam Interactions with Materials and Atoms},
  volume={355},
  pages={81--85},
  year={2015},
  publisher={Elsevier}
}

@book{madou2011fundamentals,
  title={Fundamentals of microfabrication and nanotechnology, three-volume set},
  author={Madou, Marc J},
  edition={Third},
  year={2011},
  publisher={CRC Press}
}

@article{kuiken1984etching,
  title={Etching: a two-dimensional mathematical approach},
  author={Kuiken, HK},
  journal={Proceedings of the Royal Society of London. A. Mathematical and Physical Sciences},
  volume={392},
  number={1802},
  pages={199--225},
  year={1984},
  publisher={The Royal Society London}
}

\end{document}